\documentstyle[12pt]{article}
\def\beqa{\begin{eqnarray}}
\def\eeqa{\end{eqnarray}}
\def\be{\begin{equation}}
\def\ee{\end{equation}}
\begin{document}
\begin{titlepage}
        \title{A new conformal duality of spherically 
 symmetric space--times}
\author{Hans - J\"urgen 
Schmidt$^{1}
$\thanks{http://www.physik.fu-berlin.de/\~{}hjschmi \ \ 
e-mail: hjschmi@rz.uni-potsdam.de}
\\  {\em $^1${\small Institut f\"ur Mathematik, Universit\"at
Potsdam}}\\
{\em{\small PF 601553, D-14415 Potsdam, Germany}}\\
{\em{\small and}}\\
 {\em{\small Institut f\"ur Theoretische Physik, Freie
Universit\"at Berlin,}}\\
{\em{\small Arnimallee 14, D-14195 Berlin, Germany}}\\ }
\date{}
\maketitle
\begin{abstract}
A contribution linear in $r$ to the gravitational
 potential can be created by a suitable conformal
 duality transformation: the conformal factor is $1/(1+r)^2$ 
and $r$ will be replaced by $r/(1+r)$, where $r$ is the
Schwarzschild radial coordinate.  Thus, every spherically
symmetric solution of conformal Weyl gravity is conformally
related to an Einstein space. This result finally resolves a long
controversy about this topic.    

     As a byproduct,  we present an example of a spherically
symmetric Einstein space which is a limit of a 
 sequence of Schwarzschild--de Sitter space-times but 
which fails to be expressable in Schwarzschild coordinates.
This example also resolves a long controversy. 
 \end{abstract}
\thispagestyle{empty} 
 \vspace{5.mm} Keyword(s):  Alternative Theories of
Gravity,  Conformal invariance, spherical symmetry, 
Schwarzschild coordinates, Einstein spaces.
 \vfill
\end{titlepage}

\section{Introduction}
\setcounter{equation}{0} 
From the Lagrangian 
\be
L \qquad = \qquad C_{ijkl}C^{ijkl} \quad \sqrt{-g}
\ee 
where $C^i_{\, \, jkl} $ is the conformally invariant 
Weyl tensor one gets the Bach tensor [1] 
\be
B_{ij} \qquad = \qquad 2 \ C^{k \ \ l}_{\ ij \ ;lk}
\quad + \quad C^{k \ \ l}_{\ ij } \ R_{lk}
 \ee
Recently, the solutions of the Bach equation $B_{ij}=0$, i.e.,
 the  vacuum solutions of conformal Weyl gravity, 
 enjoyed a renewed interest because in the static 
spherically symmetric case one gets a term 
linear in $r$, (cf. [2] for a deduction and for the
motivation): 
\be 
ds^2 \quad = \quad  - A(r)  dt^2 \ + \  \frac{dr^2}{A(r)}
 \ + \  r^2 d\Omega^2
\ee
with 
\be
A(r) \quad = \quad 1-3\beta \gamma 
- \frac{(2-3\beta \gamma)\beta}{r} + \gamma r - k r^2
\ee
Further dicussion of this solution can be found in [3].
In [4], the viability of the term  $\gamma r $ is
doubted, whereas in [2,3] just this part of the potential
 played the main role.\footnote{From dimensional analysis one 
can deduce the powers of $r$ in three different ways
 as follows: a) In the Newtonian limit (i.e. $\Delta$ 
is the flat-space Laplacian)
one gets from the Einstein-Hilbert Lagrangian $L_{EH}$
via $\Delta \varphi =0$ the two spherically symmetric 
 solutions $\varphi = 1$
and $\varphi = 1/r$; 
and from $L$ eq. (1) via $\Delta \Delta \varphi =0$ one
gets additionally 
 $\varphi = r$ and $\varphi = r^2$ (and, of course, all the 
linear combinations.) $\varphi = 1$ gives flat space, and
 $\varphi = r^2$ corresponds to the de Sitter space-time,
 so the essential terms are $1/r$ for  $L_{EH}$ and 
$r$ for $L$. 
 b) $L$ and $L_{EH}$
differ by a factor $<length>^2$, so this should be the case
for the potentials, too. c) Similarly one gets this as
heuristic argument by calculating the Greens functions in 
momentum space.}

\medskip

A solution of the Bach equation is called trivial if it is
conformally related to an Einstein space, i.e. to a vacuum
solution of the Einstein equation with arbitrary 
$\Lambda$ [5]. So  our question reads: Do  non-trivial 
spherically symmetric  solutions of the Bach equation exist?
Up to now, contradicting answers have been given: Metric 
(3) with (4) is an Einstein space for $\gamma = 0$ only,
 so it seems to be a non-trivial solution for 
$\gamma \ne 0$, whereas in [5] (cf. also [6] for earlier 
references) it is stated that only trivial spherically
symmetric  solutions of the Bach equation exist. 

\medskip

It is the aim of the present paper to clarify this 
contradiction by introducing a new type of conformal 
duality within spherically symmetric 
space-times.\footnote{It will be a duality
 different from that one introduced
 in [7], cf. [8] for a review on conformal transformations
between fourth-order theories of gravity.}
     The result will be that the value of $\gamma$ 
in eq. (4) can be made vanish by a conformal transformation. 
Then the question whether this linear term is physically
 measurable or not depends on the question in which of these
two  conformal frames the non-conformal matter lives. 
  
\medskip 

As a byproduct of this discussion we will present a new
 view to the question (see the different statements to this
question in [9-12]) under which circumstances a
 spherically symmetric Einstein space can  be
expressed in Schwarzschild coordinates.

\medskip

The paper is organized as follows: In sct. 2 we will deduce
the new duality transformation, in sct. 3 we apply  this
transformation to  the solution eq. (3,4), and in sct. 4
we look especially to those solutions where Schwarzschild
 coordinates do not apply.

\section{A new conformal duality transformation}

The general static spherically symmetric metric can be written

as 
\be
ds^2 \quad = \quad  - A(r)  dt^2 \  + \  B(r) dr^2 
 \ + \ C(r) d\Omega^2
\ee
where $d\Omega^2 = d\psi^2 + \sin^2 \psi d\phi^2$ is the 
metric of the standard 2-sphere. The functions $A$, $B$ and
$C$ have to be positive. 

\medskip

The main simplification for solving the Bach equation for 
 metric (5) was done in [2] as follows: The two 
 possible gauge degrees of freedom (a redefinition of the 
radial coordinate $r$ and 
 the conformal invariance of the Bach equation) can be used to
get $r$ as Schwarzschild coordinate, i.e., $C(r)=r^2$, and
$A(r) \cdot B(r) = 1$, i.e., one starts from the metric (3).
The case when Schwarzschild coordinates do not apply will
be discussed in sct. 4,  here we concentrate on the following
question: Do there exist conformal transformations of metric
(3) which keep that metric form-invariant?  

\medskip 

Of course, if $r$, $ds$, and $t$ are multiplied by the same 
non-vanishing constant $\alpha$, and the function $A$ will be
redefined accordingly, then metric (3) remains form-invariant.

This conformal transformation with a constant  conformal
factor is called a homothetic transformation, and it will not
be considered essential. Likewise, the transformation
 $r \longrightarrow -r$  not changing the form of the metric
(3) will not be considered essential. 

\medskip

Example: Let $A(r)=1-\frac{2m}{r}$, i.e., the Schwarzschild
solution with mass parameter $m$. Let $\hat r = \alpha \ r$,
 $d \hat s^2 = \alpha^2 \ ds^2$, then 
 $d \hat s^2$ represents the Schwarzschild solution 
with mass parameter $\hat m = \alpha  \ m$.\footnote{This
 applies also to negative values $\alpha$.}

\medskip

 One should expect that further conformal transformations 
do not exist because we already applied the conformal degree
of
freedom to reach the form (3) from the  form (5). This 
expectation shall be tested in the following: 

\medskip

Let $b(r)$ be any  non--constant function, and let
the conformally transformed metric be 
$d\tilde s^2 \ = \ b^2(r) \ ds^2$. With eq. (3) this reads 
\be
d \tilde s^2 \ = \  -  b^2(r) A(r)  dt^2 \ +
 \  \frac{b^2(r)dr^2}{A(r)}
 \ + \  b^2(r) r^2 d\Omega^2
\ee  
Next, we have to assume that $b(r) \cdot r$ is not
a constant, and then we can introduce $\tilde r =b(r) \cdot r$
as new Schwarzschild radial coordinate for metric (6). We get
\be
\frac{d\tilde r}{dr} \quad = \quad 
b(r) \ + \ r \frac{db}{dr}
\ee
Form-invariance in the 00-component means that  
\be
\tilde A(\tilde r) \quad = \quad b^2(r) A(r) 
\ee
and form-invariance in the 11-component implies
\be
\frac{b^2(r)dr^2}{A(r)} \quad = \quad 
\frac{d\tilde r^2}{\tilde A(\tilde r)}
\ee
Eqs. (8) and (9) together imply
\be 
\frac{d\tilde r}{dr} \quad = \quad \pm \ 
b^2(r)
\ee
If the lower sign appears we shall apply  the transformation
 $r \longrightarrow -r$ to get the upper sign. So we get 
without loss of generality from
eqs. (10) and (7) 
\be
b^2(r) \quad = \quad 
b(r) \ + \ r \frac{db}{dr}
\ee
The  non-constant solutions of eq. (11) are  
\be 
b(r) \quad = \quad \frac{1}{1+\alpha r}
\ee
with a non-vanishing constant $\alpha$.  The assumption that 
$b(r) \cdot r$ is not a constant is  always fulfilled. We get
\be 
\tilde r \quad = \quad \frac{r}{1+\alpha r}
\ee
which is valid for $1+\alpha r \ne 0$ and can be inverted to 
\be 
 r \quad = \quad \frac{\tilde r}{1+ \tilde \alpha \tilde r}
\ee
where $\tilde \alpha = - \alpha $. Eqs. (13) and (14) are 
dual to each other: Exchange of tilted and untilted 
quantities  changes the one of them to the other. 

A likewise duality can be found for eq. (8) because of
\be 
\tilde b (\tilde r) \cdot b(r) \quad \equiv \quad 1
\ee
and for eq. (12). 

\medskip

Factorizing out a suitable homothetic transformation 
we can restrict to the case $\alpha = 1$. Further, we 
restrict to the case that
 the denominator of eq. (13) is positive. Let us summarize
this restricted case as follows:

\bigskip

Let $A(r)$ be any positive function, let
$b(r) = 1/(1+ \alpha r)$ with $\alpha = 1$ and let
$$
ds^2 \quad = \quad  - A(r)  dt^2 \ + \  \frac{dr^2}{A(r)}
 \ + \  r^2 d\Omega^2
$$
Then the tilde-operator defined by $\tilde \alpha = -
\alpha$, $\tilde A(\tilde r)  =  b^2(r) A(r)$, 
$$
\tilde r \quad = \quad \frac{r}{1+\alpha r}
$$
and 
$$d\tilde s^2 \ = \ b^2(r) \ ds^2$$
represents a duality, i.e., the square of the  tilde-operator
is the identity operator.

\section{Spherical symmetry and the Bach equation} 

Let us apply the duality from sct. 2 to the 
Schwarzschild--de Sitter solution, i.e., to metric (3) with
\be 
A(r) \quad = \quad  1 \ - \ \frac{2m}{r} \ - \ 
\frac{\Lambda}{3} r^2
\ee
That means, we have to insert eq. (16) into 
eqs. (6, 12, 14). Finally, we remove all the  tildes, and
we arrive at a metric which exactly coincides with eqs. (3,4):
There is a one-to-one correspondence between the three 
parameters $m$, $\Lambda$, $\alpha$ on the one hand, and   
$\beta$, $\gamma$, $k$ on the other hand. 

\medskip

Here is the main result of the present paper: 
The  Mannheim-Kazanas [2]-solution given by 
eqs. (3,4) of the Bach equation is nothing but
a conformally transformed Schwarzschild-de Sitter metric;
the 3-parameter set of solutions (3,4) can be
found  by the 
conformal duality deduced in sct. 2.

\medskip

It should be mentioned that the set of solutions
of the full non-linear field equation is really
only 3-dimensional, and that this  is in contrast to the
linearized equation which allows all linear combinations of
1, $r$, $1/r$, and $r^2$, i.e., a 4-dimensional set.

\medskip

Up to now we had assumed that the metric is static and
spherically symmetric. However, also the Bach equation
 allows to prove  a Birkhoff-like theorem 
[6, 13]: Every spherically symmetric solution is 
conformally related to a solution possessing a fourth
isometry.\footnote{This fourth isometry may be time-like
 or space-like, and we have a regular horizon at 
surfaces where this character changes, so this
 is exactly analogous to the situation in Einstein's 
theory.}
 Another version of this result reads: 
Every spherically symmetric solution is almost everywhere 
conformally related to an Einstein space. Furthermore,
the necessary conformal factor can always be chosen such 
that it maintains the spherical symmetry.

\medskip

Why we need the restriction ``almost everywhere'' in the
second version? This applies to those points where the 
necessary conformal transformation becomes singular.
Example: Take  $u=1/r$ as
new coordinate in the Schwarzschild solution and apply an
analytic conformal transformation such that the metric can
 be analytically continued to negative values $u$ via a
regular point $u=0$; by construction, this space-time solves
the Bach equation, but at $u=0$ it fails to be conformally
related to an Einstein space.

\section{Applicability of Schwarzschild coordinates} 

To complete the discussion we want to give also those
spherically symmetric solutions of the Bach equation which
cannot be expressed in Schwarzschild coordinates. 
Before we do so, let us compare with the analogous 
situation in Einstein's theory.

\medskip

It has a long tradition to assume, see e.g. [9], that 
 every static spherically symmetric line element can be
expressed in Schwarzschild coordinates, i.e., that in metric
 (5), $C(r) = r^2$ can be achieved by a coordinate
transformation. However, the topic is a little bit more
involved:\footnote{In [10], sct. 23.2., page 595 one reads:
``For a more rigorous proof that
 in any static spherical system Schwarzschild coordinates 
can be introduced, see Box 23.3.''. But that Box 23.3. at
page 617 does not only give  this proof, but also the 
necessary assumption: `` \dots such a transformation is
 possible, (i.e. nonsingular) only where 
$(\nabla r)^2 \ne 0$.'' Later in the book (page 843) one
can find the sentence: ``The special case $(\nabla r)^2 = 0$
is treated in exercise 32.1.'' and 3 pages later ``We
thank G.F.R. Ellis for pointing out the omission of the case
$(\nabla r)^2 = 0$ in the preliminary version of this book.''
 Gaussian coordinates for metric (5), i.e., 
 $B \equiv 1$, can always be chosen by a redefinition 
of $r$, but Schwarzschild coordinates can be introduced only
 in regions where $dC/dr \ne 0$. On the other hand, 
 the Schwarzschild radius comes out after one integration
 which has the result that usually, the order of the 
field equation will be reduced  by one if 
 expressed in Schwarzschild coordinates. This latter property
is the very reason for the usefulness of them.}

\medskip

Let us take a special example of the Schwarzschild-de Sitter
metric: We insert $m=l/3 > 0 $ and $\Lambda = 1/l^2$ 
into eqs. (3,16). For any positive constant $\varepsilon$
we apply the following coordinate transformations 
\be 
r \ = \ l \ + \ \varepsilon x, \qquad \qquad
t \ = \ l^2 \tau/\varepsilon 
\ee
and get
\be 
ds^2 = - l^4 D d\tau^2  + \frac{dx^2}{D} +
(l+\varepsilon x)^2 d\Omega^2 
\ee
with
\be 
D = \frac{1}{\varepsilon^2} [1 - \frac{2l}{
3(l+\varepsilon x)} - \frac{(l+\varepsilon x)^2}{3l^2}]
\ee
Developing this $D$ in a series in $\varepsilon $ it
turns out that it is regular at $\varepsilon =0$
and there its value reads $D=-x^2/l^2$. Therefore:  Eq. (18)
represents a one-parameter family of space-times
analytic in the parameter $\varepsilon$, and for
 every  $\varepsilon > 0$ it represents a spherically
 symmetric solution of the Einstein equation with 
 $\Lambda = 1/l^2$. From continuity reasons, it represents 
a solution also for  $\varepsilon = 0$. We get
\be 
ds^2 = l^2[- \frac{dx^2}{x^2} + x^2 d\tau^2 + d\Omega^2]
\ee
which represents a spherically symmetric Einstein space 
which cannot be written in Schwarzschild coordinates. 
(It should be noted that in these coordinates, $x$ is timelike
and $\tau$ is space-like.)
The deduction of this solution presented here seems to be new.
Nevertheless, it is already known, but usually it is not 
listed within the set of spherically symmetric Einstein
 spaces: In [11] it is listed in table 10.1. under
the topic ``$G_6$ with $\Lambda$-term''. In fact, metric (20)
represents the cartesian product of two 2-spaces of constant
and equal curvature, cf. [12]. Therefore, it is also a static
metric and possesses a 6-dimensional isometry group. 

\medskip

A fortiori, metric (20) represents also a static 
spherically symmetric solution  of the Bach equation,
and this solution is not listed in refs. [2,3].

\medskip

Further, let us mention that
 the cartesian product of two 2-spaces of constant
 curvature $P$ and $Q$ resp. represents an Einstein space 
iff $P=Q$, and it represents a solution of the
Bach equation iff $P^2 = Q^2$. Thus, for $P=-Q \ne 0$
 we get another static spherically symmetric solution of the
Bach equation which cannot be expressed in Schwarzschild
coordinates; however, it is conformally flat and 
therefore trivial, too.

\medskip

Finally, we want to stress that the above consideration 
only dealt with vacuum solutions  of conformal Weyl
 gravity; of course, the inclusion of non-conformal 
matter  requests to fix one of the conformal frames, 
and it has to be discussed yet whether this shall
  be  the Schwarzschild-de Sitter or in the 
Mannheim-Kazanas frame. 

\medskip 

The result of the present paper is 
that both solutions are conformally related, and that no
further spherically symmetric solutions of the Bach equation
exist. 


\section*{Acknowledgement}
Financial support from DFG is gratefully acknowledged.
I thank the colleagues of Free University Berlin, 
where this work has been done, especially Prof.
H. Kleinert, for valuable comments.  

\bigskip

\section*{References}

\noindent
[1] R. Bach, Math. Zeitschr. {\bf 9} (1921) 110; 
 H. Weyl, Sitzber. Preuss. Akad. d. Wiss. Berlin,
 Phys.-Math. Kl. (1918) 465. 

\medskip

\noindent
[2] P. Mannheim, D. Kazanas, Gen. Relat. Grav. 
{\bf 26} (1994)  337; P. Mannheim, D. Kazanas, Phys. Rev. D
{\bf 44}  (1991) 417;  P. Mannheim, Phys. Rev. D {\bf 58} 
 (1998) 103511;  N. Spyrou, D. Kazanas, E. Esteban, Class.
Quant. Grav.     {\bf 14} (1997) 2663.

\medskip

\noindent
[3] A. Edery, M. Paranjape, Phys. Rev. D {\bf 58} (1998) 
 024011; A. Edery, M. Paranjape, Gen. Relat. Grav. {\bf 31}
 (1999) in print.

\medskip

\noindent
[4]
J. Demaret, L. Querella, C. Scheen, 
Class. Quant. Grav. {\bf 16} (1999) 749.

\medskip

\noindent
[5] H.-J. Schmidt, Ann. Phys. (Leipz.) {\bf 41} (1984) 435.

\medskip

\noindent
[6] R. Schimming, p. 39 in: M. Rainer, H.-J. Schmidt (Eds.)
Current topics in 
 mathematical cosmology, WSPC Singapore 1998.

\medskip

\noindent
[7] H.-J. Schmidt, gr-qc/9703002; Gen. Relat.
 Grav. {\bf 29} (1997) 859.

\medskip

\noindent
[8] V. Faraoni, E. Gunzig, P. Nardone: Conformal
transformations 
in classical gravitational theories and in cosmology,
gr-qc/9811047, Fund. Cosmic Physics, to appear 1999.

\medskip

\noindent
[9] M. v. Laue,  Sitzber. Preuss. Akad. d. Wiss. Berlin,
 Phys.-Math. Kl. (1923) 27.

\medskip

\noindent
[10] C. Misner, K. Thorne, J. Wheeler: Gravitation, 
Freeman, San Francisco 1973. 

\medskip

\noindent
[11] D. Kramer, H. Stephani, M. MacCallum, E. Herlt: 
 Exact solutions of Einstein's 
field equations, Verl. d. Wiss.  Berlin 1980. 

\medskip

\noindent 
[12] M. Katanaev, T. Kl\"osch, W. Kummer: Global properties 
of warped solutions in General Relativity, gr-qc/9807079.

\medskip

\noindent 
[13] H.-J. Schmidt, Grav. and Cosmol. 
{\bf 3} (1997) 185; gr-qc/9709071.       

\end{document}